# Thin glass shells for active optics for future space telescopes

G. Vecchi*, S. Basso, M. Civitani, M. Ghigo, G. Pareschi, B. Salmaso
INAF-Brera Astronomical Observatory, Via E. Bianchi 46, 23807 Merate, Italy

## ABSTRACT

We present a method for the manufacturing of thin shells of glass, which appears promising for the development of active optics for future space telescopes. The method exploits the synergy of different mature technologies, while leveraging the commercial availability of large, high-quality sheets of glass, with thickness up to few millimeters. The first step of the method foresees the pre-shaping of flat substrates of glass by replicating the accurate shape of a mold via hot slumping technology. The replication concept is advantageous for making large optics composed of many identical or similar segments. After the hot slumping, the shape error residual on the optical surface is addressed by applying a deterministic sub-aperture technology as computer-controlled bonnet polishing and/or ion beam figuring. Here we focus on the bonnet polishing case, during which the thin, deformable substrate of glass is temporary stiffened by a removable holder. In this paper, we report on the results so far achieved on a 130 mm glass shell case study.

## 1. INTRODUCTION

Space telescopes displaying better angular resolution and sensitivity than is available nowadays are required in order to advance our understanding of universe at different space and time scales. To this purpose, different strategies are under consideration and several technological issues need to be addressed. The dimensions of launch vehicle fairings limit the aperture of space telescopes. To circumvent this constraint, large space telescopes may include segmented and deployable mirrors [1]. In addition, the payload mass needs to comply with launch limitations: progress in the fabrication technologies of optical materials, in particular glass and glass ceramics, brought the areal density of optical systems to decrease, allowing larger apertures for either monolithic or segmented mirrors [2]. Different optical materials may offer favorable properties at different operating conditions. For instance, most stable optics at room temperature are usually made of glass ceramics such as Zerodur and ULE, which are characterized by extremely low coefficient of thermal expansion near room temperature. On the other hand, fused silica and borosilicate glasses may represent the preferred option for optical systems requiring stability at cryogenic temperatures, where they reach near zero thermal expansion [3]. In parallel to the developments on glass optics, new operative strategies are also conceived to overcome the restrictions owing to payload volume and mass, including concepts of on-orbit assembly of modular telescopes [4].

Concepts for very large space telescopes will require the production on a large-scale of lightweight mirrors having precisely figured surfaces. Optical demonstrators composed of thin glass substrates, i.e., shells, coupled to rigid supporting structures were considered to reduce the mass, while keeping the required stiffness [3, 5]. A discrete set of adjustable actuators allows compensating for the in-flight deformations induced on the glass shells. Moreover, active optics helps to loose the stringent requirement on the optical manufacturing and the system performance stability [6, 7].

Methods for manufacturing thin glass optics were developed, often in the framework of adaptive optics solutions [8], which improve the performance of present ground-based telescopes. The manufacturing processes may involve the thinning and grinding of high-quality thick blanks in combination with stress polishing [9], stressed lap figuring [10], polishing and figuring [11-13] technologies. The baseline process of thinning a high value blank is a delicate step so far successfully applied for the production of monolithic units or few segments.

Cost and time of manufacturing are expected to be relevant aspects for the production of several units or segments required for the assembly of modules composing large segmented mirrors. Technological advancements on mirror replication processes were reported [14, 15], which allowed reaching the required optical tolerances by eliminating or limiting the costly grinding and polishing processes.

---

*gabriele.vecchi@inaf.it; phone +39 02 72320478; fax +39-02 72320601; www.brera.inaf.it

In this paper, we present a concept to manufacture thin glass shells. The process starts with the procurement of flat sheets of high-quality glass commercially available off-the-shelf. In particular, borosilicate glasses can leverage on the continuous development driven by the global market of flat panel display technologies. Thanks to it, large and thin substrates of borosilicate glass are widely available with thickness in the millimeter range, tight uniformity of thickness down to few tens of microns, and low roughness values ≤ 1 nm Root Mean Square (RMS).

In its first phase, the process provides the pre-shaping of the glass foils by hot slumping technology, based on the replication of the surface of a master mold. In the second phase of the process, focus of the present work, the residual surface error after slumping is addressed by high-precision sub-aperture polishing and figuring technologies. The flow chart in Fig. 1 indicates the key steps of the manufacturing approach. The innovative aspect relies on the application of polishing/figuring on the surface of the shell pre-shaped by slumping, while it is temporary stiffened on a blocking substrate. Leveraging on the replica concept and avoiding the thinning and grinding steps from high quality blanks, the proposed approach is promising to reduce the time and cost of fabrication, particularly for the serial production of many segments for large mirrors.

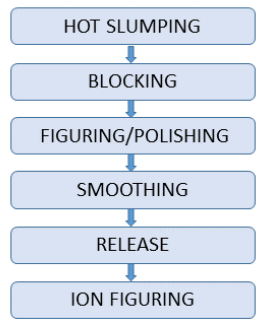

Figure 1. Manufacturing steps for thin glass shells. The shell is blocked on a stiffening support before the polishing/smoothing activity and released afterwards.

We previously conducted preliminary experiment of bonnet polishing on 1-mm thin fused silica wafers 100 mm across [16]. The present work is the follow up of the test activity by bonnet polishing conducted on a small-scale 130 mm slumped glass shell [17]. In Section 2, we describe metrology and bonnet polishing results obtained on the same 130 mm sample after improvement of the temporary blocking process of the flexible shell onto a stiff holder. In Section 3, we summarize the conclusions and the foreseen future activity.

### 1.1 Hot slumping technology

The technological development of the hot slumping of glass performed at INAF-Brera Astronomical Observatory started in the framework of E-ELT Design Study supported by the European Community under OPTICON-FP6 [18]. The aim was to develop a process suited for the production of a large number of thin glass segments for adaptive optics. The hot slumping process employs an oven, where a thin, flat foil of glass is placed on top of a master mold. The surface of the mold is previously figured to meet the target shape with high accuracy. After application of a suitable thermal cycle, the heated glass softens and slumps onto the mold surface, copying its shape. After cooling down of the shell-mold system, the slumped glass shell is released from the mold. The results were encouraging but displaying an error still too high compared to the requirements: we achieved a surface RMS error >100 nm versus a requirement <10 nm over 100 mm spatial scale [19]. In the present study, we focus on a simple case, namely, spherical shells made in BOROFLOAT® 33 glass, 130 mm wide and 2 mm thick. These shells, together with larger spherical shells up to 500 mm [20], were slumped to replicate the shape of a spherical mold in the early stage of the hot slumping development carried out at INAF-Brera Astronomical Observatory.

### 1.2 Deterministic polishing and figuring technology

We plan to execute the post-slumping correction of the optical surface by deterministic sub-aperture figuring/polishing technologies. Once the optical surface has been polished and figured to meet the requirements, the thin shell of glass (featuring an areal density <5 Kg/m$^2$ in the case under study) can be integrated into a lightweight mechanical assembly, leading to overall values of areal density <20 Kg/m$^2$ [6, 21]. In this study, we investigate the performance of bonnet polishing technology, while the behavior of Ion Beam Figuring (IBF) either on thin slumped shells directly [22] or after bonnet polishing will be a matter for future activity.

An IRP (Intelligent Robotic Polisher) 1200 series machine by Zeeko Ltd. is implemented at INAF-Brera Astronomical Observatory [23, 24]. The machine relies on a sub-aperture rubber tool, the bonnet, to polish the surface and correct the error shape while scanning through it along a defined path. The bonnet is inflated by air pressure and is covered with a polishing medium such as polyurethane foam. Dedicated nozzles deliver a temperature-controlled and density-monitored abrasive slurry to the workpiece-tool contact area. The material removal rate is proportional to the relative surface speed and to the pressure applied to the workpiece, according to the Preston model [25]. Since surface speed and pressure are usually set constant into a run, the local removal is proportional to the dwell time of the tool at any position. A dwell time matrix is calculated in order to correct the measured error map of the surface. The machine executes the dwell time matrix by varying the tool speed along the path. The smallest bonnet available is suited to address error with spatial wavelength down to the 5-10 mm range, while different rigid tools are required to smooth ripples of higher frequency more efficiently, e.g., in the 0.5-5 mm range.

IBF technology [26-28] is also well suited for the correction of low frequency errors, employing the dwell time principle to remove material, similarly to the bonnet polishing technology. On the other hand, featuring as a contactless tool the sub-aperture beam of accelerated ions hitting against the optical surface, IBF enables the correction of print-through and deformations induced on thin and lightweight optics by previous manufacturing steps and/or due to supporting mounts. Similar advantage applies to correct the near edge shape of optics, as in segmented mirrors.

## 2. THIN GLASS SHELLS

### 2.1 Metrology of the thin glass shell

Thin optics like the shells of glass investigated here tend to sag due to gravity, and to deform depending on the supporting mount. We performed Finite Element Method (FEM) simulation to minimize the deformation induced on the optical surface of the shell during the interferometric measurements. The mounting system used for the interferometric measurements is shown in Fig.2A. It is a non-kinematic 3-points support, where the third point on the top acts like a stopper to prevent the glass sample from falling backwards. According to FEM results, in order to minimize the deformation on the optical surface, we tilted backwards the glass shell by about 1.5 mm [17].

We used a Zygo GPI series Fizeau interferometer equipped with a 4-inch f/3.3 transmission sphere to measure the form error map of thin shells. BOROFLOAT® 33 glass shells were made from circular disks of 130 mm diameter and 2 mm thickness, and they were slumped to a spherical shape with radius of curvature of about 4 m. The back surface of the shell is rough to avoid perturbing the measurements of the front surface. We set a relay mirror to fold the interferometric setup and fit within the optical bench. In the previous work, we employed such setup and the support in Fig.2A to monitor the evolution of the surface error map through a few iterations of the bonnet polishing process. At the beginning of the present study, we realigned the entire setup *ex novo* to repeat the measurements of the optical surface with the shell in free-standing mode, supported as described above. Fig.2B shows the measured residual error map after removal of tilt and power terms. The map refers to a clear aperture set to 110 mm, and displays an RMS value of 30 nm, and Peak to Valley (PV) of 170 nm. We compared this result to measurements performed in free-standing mode previously, at the end of the first iterative test of bonnet polishing [17], when we had obtained a map with 38 nm RMS error. After subtraction of the averaged error maps measured in the two distinct sessions, we found the repeatability error map shown in Fig.2C, featuring an astigmatic shape of about 10 nm RMS. We suppose such a change of shape on the optical surface arises because of the non-kinematic mount, as a different amount of stress applies on the flexible shell at each repositioning. The surface and repeatability error maps appear also perturbed by both dust entrapped in the optical path during metrology and by local defects, probably generated on the surface by dust during the hot slumping process.

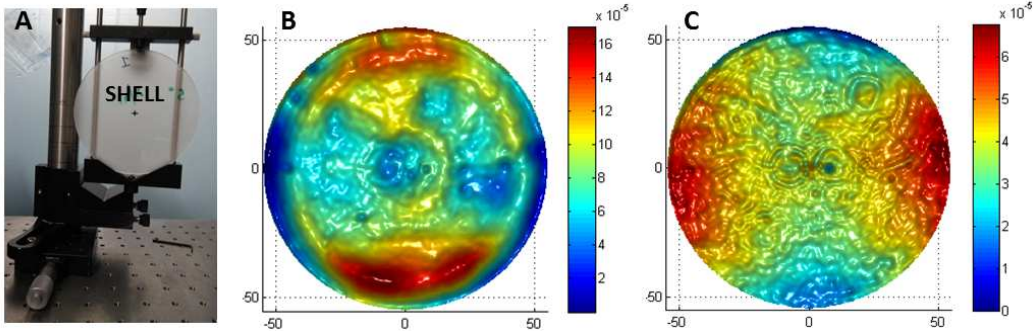

Figure 2. A. Slumped shell 2-mm thick supported on a non-kinematic 3-points mount. B. Residual error map of the shell over 110 mm clear aperture, 30 nm RMS and 170 nm PV error after removal of tilt and power terms. C. Repeatability error map, 10 nm RMS and 68 nm PV. Color scales are in millimeters.

**2.2 Temporary stiffening of the thin glass shell**

Thin substrates of glass are flexible and need to be stiffened temporarily during the phase of bonnet polishing. Otherwise, the pressure applied within the contact area on the shell's optical surface would distort its local shape, perturbing the dwell-time-based process of material removal. A stiffening holder was designed as mechanical flange for interfacing the glass shell both to the polishing machine and to the interferometric setup. In fact, we aimed to measure the surface error after each iteration of bonnet polishing, yet avoiding releasing the glass shell from the holder after every run. The findings of previous activity confirmed the obvious fact that the ideal stiffening element and blocking process should induce the least deformation to the optical surface, as it would in free-standing mode. Therefore, before performing a new polishing iteration, we re-machined the stiffening flange to achieve a better matching to the shell's shape. However, we did not meet the 4 m target curvature, estimating a radius near 4.7 m instead.

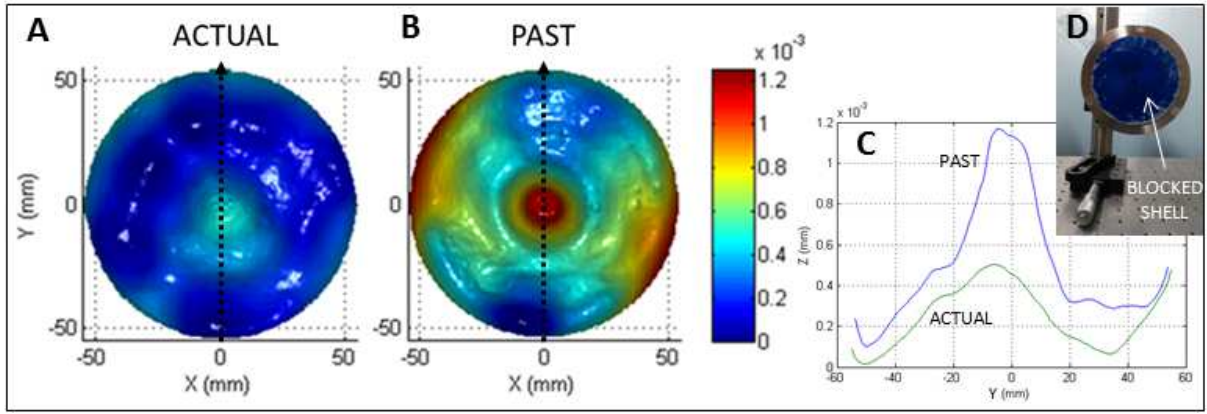

Figure 3. A. Error map of the shell blocked on the stiffening flange, over 110 mm clear aperture, after removal of tilt and power terms. B. Error map of the blocked shell as measured in the previous activity [17]. C. Profiles along y-axis extracted from panels A (actual) and B (past). D. Picture of the shell blocked onto the stainless steel flange using a (blue) thermal wax. The shell-flange system is mounted in the measuring setup. Color scale in millimeters is common to panels A and B for sake of comparison.

Fig.3A shows the error map of the shell stiffened on the holder, measured with the interferometer over a 110 mm clear aperture, after removal of tilt and power terms. For sake of comparison, we show in Fig.3B the error map of the shell stiffened on the holder as it was measured at the end of the previous polishing cycle. Fig.3C displays two profiles extracted from the two maps. The value of RMS (PV) error was reduced from previous 221 (1241 nm) to actual 106 nm (527 nm). Although we cannot directly infer that the shape improvement of the re-machined holder caused the reduction of deformation of the blocked shell, we believe it did contribute to the result. We also remark that the error map of the blocked shell (106 nm RMS) is much larger than its free standing map (30 nm RMS). Indeed, the present shell-flange system, shown in Fig.3D, still displays a thermo-mechanical mismatch between the Inox flange and the BOROFLOAT® 33 glass slumped shell. Despite it, as we will discuss in the next paragraph, we were able to correct the low frequency error further.

## 2.3 Bonnet polishing of the thin glass shell

We applied the bonnet polishing process to the thin shell blocked onto the holder. The sub-aperture polishing was set to address the low frequency content, with spatial wavelength >10 mm. Fig.4 displays three error maps obtained by measuring the shell in free standing mode throughout the corrective process. The 128 nm RMS error map (panel A) was measured at the beginning of the activity, before polishing [17]. In the present work, the starting error map is the endpoint of prior activity, shown in Fig. 4B, featuring 30 nm RMS and 170 nm PV error. We therefore performed two iterations, namely, run4 and run5, and we detached the shell from the Inox holder only after run5. Afterwards, we measured the free standing map shown in Fig. 4C, which displays an halved RMS error of 15 nm (~$\lambda/42$ with $\lambda$=633 nm) and 105 nm PV (~$\lambda/6$), over 110 mm aperture and after removal of tilt and power terms. The groove crossing the run5 map vertically occurred during run4, which was executed with the rastering direction of the bonnet along y-axis. The input correction in run5 mitigated the depth of the groove though without eliminating it. The groove is believed to have been caused by a perturbation occurred to the polishing process; it might be a sudden change of a parameter (e.g., the density of the abrasive fluid) or a local irregularity on the surface intersecting the tool path. The surface of the shell contains point-like defects, also visible in the measured maps, which we suppose likely related to the presence of dust grains at the mold-glass interface during the hot slumping process. Regardless their origin, the correction of these mid frequency features requires a smoothing process.

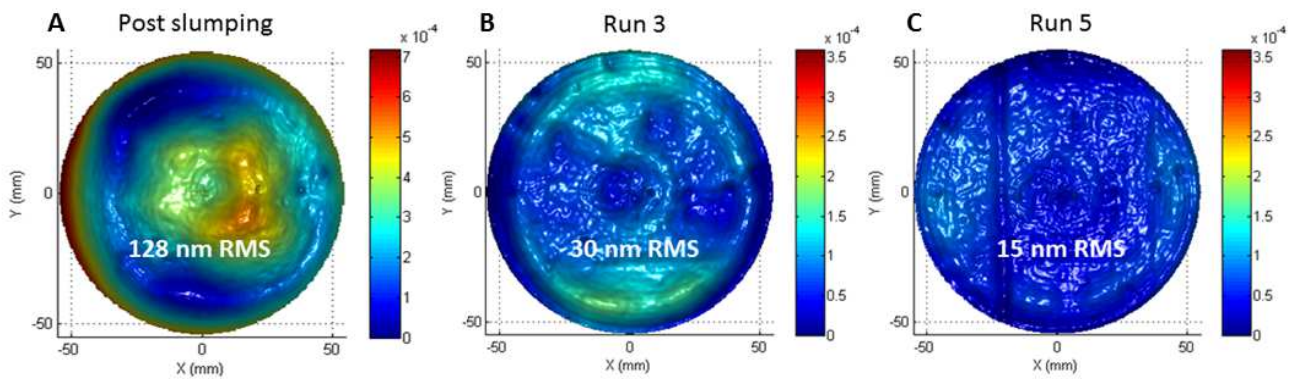

Figure 4. Trend of the free standing error map through the entire polishing process. A. Error map of the thin shell after hot slumping. B. Residual error after third iteration of polishing and starting point of the present work. C. Residual error after fifth iteration. For each map, the clear aperture is 110 mm, tilt and power terms are removed, and the value of RMS error is displayed. Color scales are in millimeters and, for sake of clarity, the amplitude range in panel A is twice that in B and C.

The value of surface RMS error achieved after bonnet polishing is close to the figuring specification expected for a glass shell before any error compensation by the actuators (flattening) in the framework of an adaptive/active system [19]. We achieved the 15 nm RMS error by polishing the thin shell glued on the stiffening holder, where it was displaying the surface error map of Fig.3A with initial RMS (PV) error of 106 nm (527 nm). The same map is shown again in Fig.5A to the purpose of comparing it with the final free standing error map, shown in Fig.5B. In this last figure, the error amplitude is multiplied by a factor 5, to enhance the comparison and highlight that no significant print through occurred during the last two runs of polishing (run4 and run5). We plot in Fig.5C two profiles cut along the y-axis of the measured maps, the surface error of the shell either blocked (Fig.5A) or free standing (cut as shown in Fig.5B but with plotted data not corrected by the multiplication factor 5).

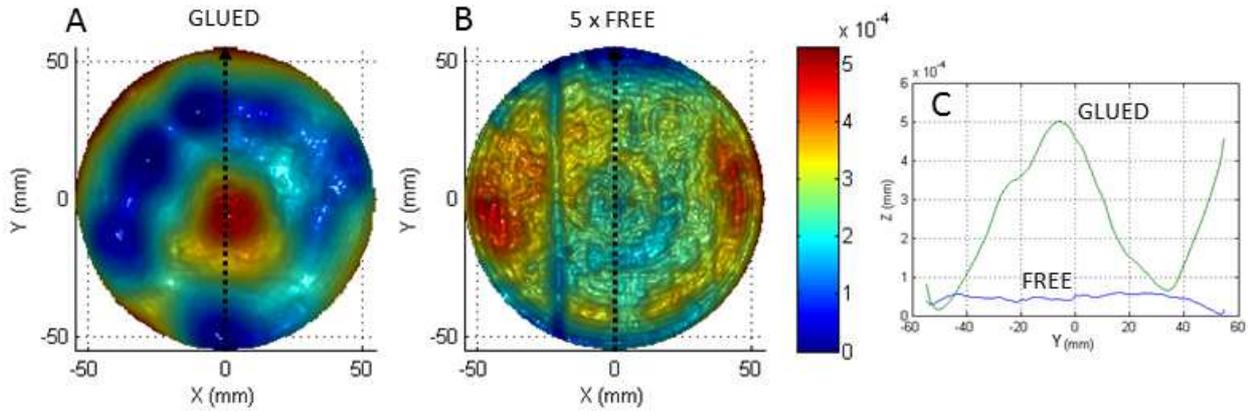

Figure 5. A. Error map of the shell glued on the stiffening holder. Same map as in Fig.3A but with amplitude scale fitting the PV value of 527 nm. B. Free standing error map after fifth iteration of bonnet polishing, with amplitude scale multiplied by factor 5 to enhance the comparison with the map in panel A. Color scale is in millimeters and common for A, B panels. For each map, the clear aperture is 110 mm, tilt and power terms are removed. C. Profiles along y-axis extracted from the error map of the shell measured in glued and free standing mode.

The repeatability RMS error of separate series of measurements of the shell blocked on the holder was found in the 10-15 nm range. This is comparable with the surface accuracy of the shell measured in free standing mode. Therefore, since the input error map feeding the next polishing iteration is derived from the subtraction of successive measurements of the shell in the blocked mode, it turns out that we reached the repeatability level of the used interferometric setup. Fig.6A shows the repeatability error obtained by subtracting two series of measurements of the blocked shell. Fig.6B is the residual error map of the shell. As clearly visible from the colored scale and from the profiles plotted in Fig.6C, the repeatability and surface maps have error amplitudes of the same order of magnitude.

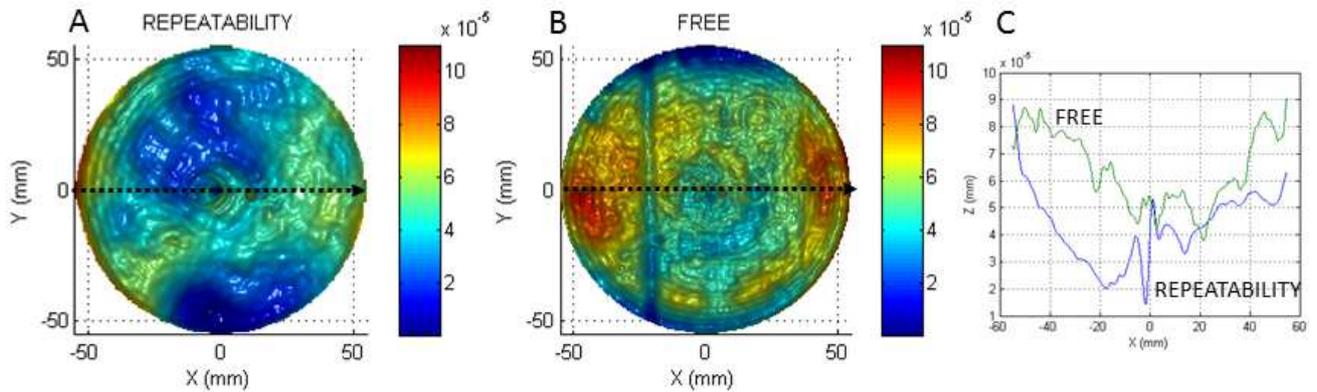

Figure 6. A. Repeatability of separate (timeframe of days) series of measurements of the shell glued on the stiffening holder. B. Free standing error map after fifth iteration of bonnet polishing. Color scale is in millimeters and common to A, B panels. C. Profiles along x-axis extracted from the maps shown in panels A and B.

As already remarked for the maps in Fig.2, we highlight that mid-high frequency undulations affecting the maps of Fig.6 are often metrological artifacts. However, some of the features visible in Fig. 6B, refer to polishing-induced waviness left on the shell's optical surface, as the dimple near x=-20mm and the ~20 nm high jump at x=40 mm in the profile of the free standing map shown in Fig. 6C.

We monitored the micro-roughness of the optical surface before and after the bonnet polishing, although we did not optimize the process for reducing the high spatial frequency content. Fig.7A shows a picture of the Micro-Finish Topographer (MFT) placed on top of the glass shell to measure the micro-roughness in the central area of the surface [29]. We equipped the MFT with an interference microscope objective enabling the measurement over a ~1mm$^2$ field of view. Figures 7B and 7C highlight the different surface pattern before and after bonnet polishing, respectively. RMS error did not change much, ranging near 2.5-3 nm. PV error appeared more erratic from one measurement to another, displaying much reduced values after polishing, down to a few tens of nanometers in the best cases. The relatively high PV values of micro-roughness in Fig.7B is related to the not-yet-optimized process of hot slumping in its early phase of development, when the shell under study was formed. The micro-roughness is expected to improve further with bonnet polishing by changing the average size of abrasive particles.

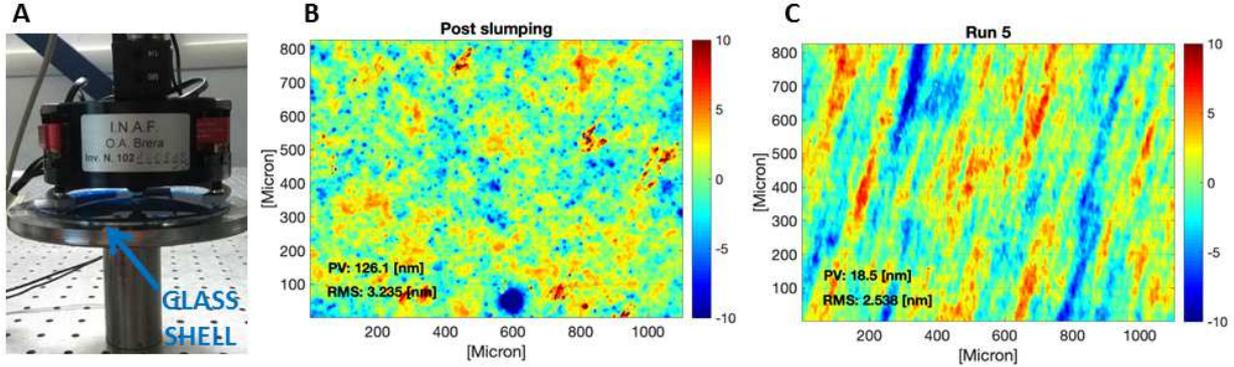

Figure 7. A. Micro-Finish Topographer measuring the thin shell blocked on the stiffening flange. B. High frequency error map measured at one central position of the surface before bonnet polishing (error values ~3.2 nm RMS, 126 nm PV). C. High frequency error map measured at one central position of the surface after bonnet polishing (error values ~2.5 nm RMS, 18.5 nm PV). Color scale is in nanometers and common to B and C panels.

## 3. CONCLUSIONS

In this paper, we presented a method for manufacturing high-precision thin glass substrates. Thin glass shells are key elements driving the development of both adaptive optics solutions in ground-based telescopes and active optics concepts for future large space telescopes. We aim to demonstrate the feasibility of the presented manufacturing approach, whose innovative aspect stands on the application of polishing/figuring on the thin glass shell pre-shaped by slumping. We highlight that it relies on cost-effective solutions for glass procurement and mirror segments replication. Large sheets of glass with tight thickness control and low micro-roughness are currently available. Then, a replication mold is used to thermally-shape the thin glass substrates by hot slumping technology. The following sub-aperture polishing/figuring process applies on the thin slumped shells to correct the residual surface errors. Finally, the shells can be assembled into lightweight supporting structures, targeting overall values of areal density <20 Kg/m$^2$ [6, 21].

We considered glass substrates 2 mm thick and with 130 mm diameter, and we focused on the bonnet-polishing as a technology to improve the shape accuracy of slumped thin optics. Due to the force applied on the shell surface during bonnet polishing, the shell needs to be blocked temporarily on a rigid support. Although we improved the blocking process of the shell as compared with previous test [17], a deformation was still induced on the optical surface. Despite it, we reached a shape accuracy of 15 nm RMS (~$\lambda$/42) and 105 nm PV (~$\lambda$/6), over 110 mm clear aperture and after removal of tilt and power terms. This value of surface RMS error is close to the figuring specification expected for a glass shell before active compensation foreseen in adaptive/active systems [19].

In the next activity we aim in particular to: 1) test the IBF process on thin slumped shells with or without previous application of the bonnet polishing, and 2) test solutions to improve the smoothing of the ripples with ~0.5-5 mm spatial wavelength, a range where the convergence factor of bonnet polishing and IBF correction is very poor. In this study, the size of the optics was limited to 130 mm (110 mm aperture considered). Our purpose is to explore the feasibility of the manufacturing process by upgrading it to a 500 mm size scale.